\documentclass{ws-procs975x65}

\begin{document}

\title{LOOP QUANTUM GRAVITY AND THE CMB:\\ TOWARD PRE-BIG BOUNCE COSMOLOGY}

\author{Aurelien Barrau}

\address{Laboratoire de Phyique Subatomique et de Cosmologie, UJF-CNRS-IN2P3-INPG\\
53, avenue des Martyrs, 38026 Grenoble cedex, France\\
E-mail: Aurelien.Barrau@cern.ch}

\begin{abstract}
This brief article sums up the possible imprints of loop quantum gravity effects on the cosmological microwave background.
We focus on semi-classical terms and show that "Big Bounce" corrections, together with the
"pre Big Bounce" state, could modify the observed spectrum.
\end{abstract}

%\keywords{Loop Quantum Gravity, Loop Quantum Cosmology}

\bodymatter
\bigskip

Loop Quantum Gravity (LQG) is a promising attempt to provide a non-perturbative background-independent quantization
of General Relativity (see, {\it e.g.}, $[$\refcite{rovelli1}$]$). Although a fully rigorous treatment of cosmological
solutions within the LQG framework is still missing, Loop Quantum Cosmology (LQC) inherits certain key physical
aspects of the full theory (see, {\it e.g.}, $[$\refcite{bojo0}$]$). As the direct measurement of the discreteness of 
space is far beyond our experimental possibilities and as the violation of Lorentz invariance is highly controversial
and by no means a {\it necessary} consequence of quantum gravity (see, {\it e.g.}, $[$\refcite{lorentz}$]$), cosmology
could very well be the best -- if not only -- observational access to Planck scale physics.  
Among the most important results of LGC, one can 
cite : the correct infra-red behavior of the theory, the resolution of the Big Bang singularity through the emergence
of a "bounce" due to repulsive quantum geometry effects at the Planck scale, the existence of a maximum density which
does not depend on the details of the model, and the existence of a minimum volume for the Universe which scales linearly with
$p(\phi)$ and can be much higher that the Planck volume (see, {\it e.g.}, $[$\refcite{ash}$]$ and references therein).\\

At the semi-classical level, the most striking result of LQC is that, in the context of a universe sourced by a
minimally coupled scalar field, the corrections cause an antrifrictional effect which accelerates the field along its
self-interaction potential. This effect can push the field up its potential and set the initial conditions for 
subsequent slow-roll inflation (see, {\it e.g.}, $[$\refcite{infla}$]$). Important features on the CMB were derived in
$[$\refcite{tsu}$]$. Interestingly, LQC corrections also generically
lead to a period of superinflation during which the Hubble factor rapidly increases rather than remaining nearly constant
as in usual inflation (see, {\it e.g.}, $[$\refcite{superinfla}$]$).\\

The main loopy corrections are the so-called "inverse volume" and  "holonomy" terms. The first ones arise for inverse powers of the densitized
triad, which when quantized becomes an operator with zero in the discrete part of its spectrum thus lacking a direct inverse. The
second ones are due to the fact that loop quantization is based on holonomies, {\it i.e.} exponentials of the connection rather
than direct connection components.\\

Corrections to gravitational wave dispersion were derived in $[$\refcite{bojo1}$]$ and quite a lot of works have already been
devoted to gravity waves in LQC (in addition to the previously cited articles, see, {\it e.g.}, $[$\refcite{lqcgen}$]$).
Recently, we have focused on a rigorous derivation of the primordial tensor power spectrum under the assumption of a
standard inflationary evolution of the background. This picture is both heuristically justified to decouple the effects
and physically meaningful as superinflation alone cannot cure all the cosmological problems. We have shown that both
holonomy (see $[$\refcite{grainlqg2}$]$) and inverse volume (see $[$\refcite{grainlqg3}$]$) corrections could lead
to observational footprints in the B-mode of the CMB. In the inverse volume case, the power spectrum should even
exhibit an exponential divergence in the IR limit with a very strong running of the index.\\

The next logical step is to include simultaneously holonomy and inverse volume terms. We have shown (in $[$\refcite{grainlqg4}$]$) the resulting  
effective gravitational hamiltonian to be, with 
the usual LQC notations:
\begin{eqnarray*} \label{Hgravpertcor}
H = \frac{1}{2 \kappa}\int_{\Sigma}\mathrm{d}^3x \bar{N} S(\bar{p},\delta E^a_i)
\left[-6\sqrt{\bar{p}}
\left(\frac{\sin\bar{\mu}\gamma\bar{k}}{\bar{\mu}\gamma}\right)^2
- \frac{1}{2\bar{p}^{3/2}} 
\left(\frac{\sin\bar{\mu}\gamma\bar{k}}{\bar{\mu}\gamma}\right)^2
(\delta E^c_j\delta E^d_k\delta_c^k\delta_d^j) \right. \nonumber\\
+ \left.\sqrt{\bar{p}} (\delta K_c^j\delta K_d^k\delta^c_k\delta^d_j) 
- \frac{2}{\sqrt{\bar{p}}} 
\left(\frac{\sin 2\bar{\mu}\gamma\bar{k}}{2\bar{\mu}\gamma}\right)
(\delta E^c_j\delta K_c^j) 
- \frac{1}{\bar{p}^{3/2}} (\delta_{cd} \delta^{jk}  E^c_j \delta^{ef} 
\partial_e \partial_f  E^d_k ) 
\right].
\end{eqnarray*}
With an appropriate change of variables, the associated equation of motion can be translated in a Sch\"odinger-like 
equation and analytically solved for some specific parameters. The resulting spectrum still exhibits the 
characteristic LQC features, dominated by the inverse volume correction. This is to be contrasted with the quantum Friedmann
equation
$
\mathcal{H}^2 = a^2 \frac{\kappa}{3} \rho \left(S- \frac{\rho}{\rho_c} \right),
$
independantly obtained in $[$\refcite{calcagni}$]$, where the holonomy correction (associated with the $-\rho^2$ term) dominates over
the inverse volume correction (associated with the $S$ term) for large values of $\rho$.\\

On the other hand, it is also important to take into account the correct ({\it i.e.} quantum-modified) background dynamics.
A realistic model was built in $[$\refcite{jakub}$]$ where the bounce naturally sets the initial conditions for inflation.
In such approaches, the resulting spectrum depends not only on the mathematical structure of the LQC corrections but also on
the choice of the field configuration {\it before} the bounce, therefore opening a pre Big Bounce window on the universe.
Generically, the power spectrum is suppressed for the largest scales and exhibits a few oscillations before recovering the
standard GR behavior in the UV limit. This study however neglects corrections associated with the mode propagation and accounts
for holonomy corrections only.\\ 

Loop Quantum Cosmology is now a field of research mature enough to produce reliable quantitative estimates that could allow for a
clear test of the theory, in particular through
modifications if the B-mode CMB power spectrum. In the forthcoming years, several important points are to be investigated:
\begin{itemize}
\item Model $[$\refcite{jakub}$]$ should be extended to account for corrections on the mode propagation.
\item The status of inverse volume correction (which is somehow controversial) should be clarified
$[$\refcite{ashperso}$]$.
\item Following $[$\refcite{bojo4}$]$, not only inverse volume but also holonomy corrections should be computed for the scalar
modes.
\item The predictions of LQC are to be compared with those of other alternative theories, in particular string cosmology.
\end{itemize}

It should also be mentioned that the source term, which has often been forgotten or miscomputed, has to be correctly
included in the analysis $[$\refcite{tom}$]$.

First attempts already show that the LQC-modified primordial power spectrum, when used as an input to compute the resulting
temperature and polarization CMB spectra, lead to substantial deviations.

\bibliographystyle{ws-procs975x65}
\bibliography{ws-pro-sample}

\begin{thebibliography}{9}
\bibitem{rovelli1} C. Rovelli, {\it  Quantum Gravity}, Cambridge, Cambridge University Press, 2004; C. Rovelli, Living
Rev. Relativity, {\bf 1}, 1 (1998); L. Smolin, arXiv:hep-th/0408048v3; T. Thiemann, Lect. Notes Phys. {\bf 631}, 41
(2003); A. Perez, arXiv:gr-qc/0409061v3.
\bibitem{bojo0} M. Bojowald, Liv. Rev. Rel. {\bf 11}, 4 (2008); A. Ashtekar, Gen. Rel. Grav. {\bf 41}, 707 (2009)
\bibitem{lorentz} C. Rovelli and S. Speziale, Phys.\ Rev.\ D {\bf 67}, 064019 (2003)
\bibitem{ash} A. Ashtekar, J. Phys. Conf. Ser. {\bf 189}, 012003 (2009). 
\bibitem{infla}  M.~Bojowald, J.~E.~Lidsey, D.~J.~Mulryne, P.~Singh and R.~Tavakol,
  Phys.\ Rev.\ D {\bf 70}, 043530 (2004);
  J.~E.~Lidsey, D.~J.~Mulryne, N.~J.~Nunes and R.~Tavakol,
  Phys.\ Rev.\ D {\bf 70}, 063521 (2004);
  D.~J.~Mulryne, N.~J.~Nunes, R.~Tavakol and J.~E.~Lidsey,
  Int.\ J.\ Mod.\ Phys.\ A {\bf 20}, 2347 (2005);
  D.~J.~Mulryne, R.~Tavakol, J.~E.~Lidsey and G.~F.~R.~Ellis,
  Phys.\ Rev.\ D {\bf 71}, 123512 (2005);
  N.~J.~Nunes,
  Phys.\ Rev.\ D {\bf 72}, 103510 (2005);
     S. Tsujikawa, P. Singh and R.Maartens,
Class. Quant. Grav. {\bf 21}, 5767 (2004).
\bibitem{tsu} S. Tsujikawa, P. Singh, and Roy Maartens, Class. Quant. Grav. {\bf 21} 5767 (2004). 
\bibitem{superinfla} 
D.~J.~Mulryne and N.~J.~Nunes, Phys.\ Rev.\ D {\bf 74}, 083507 (2006); E.~J.~Copeland, D.~J.~Mulryne, N.~J.~Nunes and M.~Shaeri,
Phys.\ Rev.\ D {\bf 77}, 023510 (2008). 
\bibitem{bojo1} M. Bojowald and G.M. Hossain, Phys. Rev. D {\bf 77}, 023508 (2008).
\bibitem{lqcgen} E.J. Copeland, D.J. Mulryne, N.J. Nunes, and M. Shaeri, Phys. Rev. D {\bf 79}, 
023508 (2009); J. Mielczarek and M. Szydlowski, Phys. Lett. B {\bf 657}, 20 (2007); 
J. Mielczarek, J. Cosmo. Astropart. Phys. 0811:011 (2008); J. Mielczarek, Phys. Rev .D {\bf 79}, 123520
(2009); M. Shimano and T. Harada, arXiv:0909.0334v2 [gr-qc].
\bibitem{grainlqg2} J. Grain and A. Barrau, Phys. Rev. Lett {\bf 102}, 081301 (2009); A. Barrau and J.
Grain, arXiv:0805.0356v1 [gr-qc]; J. Grain, arXiv:0911.1625v1 [gr-qc]
\bibitem{grainlqg3} J. Grain, A. Barrau, A. Gorecki, Phys. Rev. D {\bf 79}, 084015 (2009).
\bibitem{grainlqg4} J. Grain, T. Cailleteau, A. Barrau and A. Gorecki, arXiv:0910.2892 [gr-qc].
\bibitem{calcagni} G. Calcagni, G.M. Hossain, Adv. Sci. Lett. {\bf 2}, 184 (2009). 
\bibitem{jakub} K. Mielczarek, arXiv:0908.4329v1 [gr-qc].
\bibitem{ashperso} A. Ashtekar, private communication
\bibitem{bojo4} M. Bojowald, G.~M. Hossain, M. Kagan, and S. Shankaranarayanan, Phys. Rev. D {\bf 79} 043505 (2009). 
\bibitem{tom} see the contribution of T. Cailleteau in this volume.
\end{thebibliography}

\end{document}